\documentclass[a4paper]{jpconf}
\usepackage{graphicx}
\begin{document}
\title{Relation between high-energy quasiparticles of quasi-one-dimensional antiferromagnets in a magnetic field and a doublon of a Hubbard chain}

\author{Masanori Kohno}

\address{International Center for Materials Nanoarchitectonics (MANA), 
National Institute for Materials Science (NIMS), Tsukuba 305-0044, Japan}

\ead{KOHNO.Masanori@nims.go.jp}

\begin{abstract}
In spin-1/2 one-dimensional Heisenberg antiferromagnets and anisotropic triangular Heisenberg antiferromagnets, 
high-energy states carrying considerable spectral weights have been observed in a magnetic field using inelastic neutron scattering. 
Such high-energy properties cannot be explained in terms of either Nambu-Goldstone bosons due to spontaneous breaking of continuous symmetries 
or quasiparticles in a Tomonaga-Luttinger liquid. 
Here, we show that the mechanism causing the high-energy states is analogous to that of the upper Hubbard band 
in the one-dimensional Hubbard model, by theoretically tracing the origin of the high-energy states back to string solutions of the Bethe ansatz. 
\end{abstract}

\section{Introduction}
In conventional magnets with magnetic long-range order, dynamical properties are explained in terms of magnons in spin-wave theory \cite{SWF,SWAF}. 
In a magnetic field, the dispersion relation of the magnon deforms according to the change of the ground-state spin configuration \cite{SWH}. 
However, for systems without magnetic long-range order, the spin-wave theory based on classical spin configurations is not justified, and 
the dynamical properties are not necessarily explained in terms of the magnon. 
In fact, in a one-dimensional (1D) spin-1/2 antiferromagnetic Heisenberg chain, whose ground state has no magnetic long-range order, 
the dynamical properties in zero field are explained in terms of spinons \cite{FT,HSspinon}. 
Also, in spatially anisotropic two-dimensional (2D) frustrated Heisenberg antiferromagnets, bound states of spinons, called triplons, dominate the spectra in zero field \cite{Kohno2DHeis}. 
Although the triplon might appear similar to the magnon, its stabilization mechanism differs from that of the magnon: 
the triplon is induced by spin exchange, i.e., two-spinon hopping, between neighboring chains \cite{Kohno2DHeis}, 
while the magnon is stabilized by the classical spin configuration \cite{SWF,SWAF}. 
\par
The difference between such quasiparticles (QP's) without ordered ground states and the magnon created from magnetic long-range order becomes clearer in a magnetic field. 
For example, in an antiferromagnetic Heisenberg chain in a magnetic field, low-energy properties are primarily explained in terms of QP's called psinons ($\psi$'s) 
and anti-psinons ($\psi^*$'s) \cite{KarbachPsinon,Kohno1DHeisH}, 
which have gapless points at incommensurate momenta different from those of the magnon, in addition to the gapless points at commensurate momenta. 
In anisotropic-2D frustrated Heisenberg antiferromagnets, bound states of $\psi$'s and $\psi^*$'s dominate the low-energy spectra, 
leading to incommensurate magnetic long-range order whose momentum shifts significantly with the magnetic field and multi-particle crossover in a magnetic field \cite{Kohno2DHeisH}. 
These features cannot be explained in terms of the magnon in linear spin-wave theory. 
\par
Furthermore, in an antiferromagnetic Heisenberg chain, a continuum is separated from other low-energy continua in a magnetic field, 
and shifts to higher energies as the magnetic field increases \cite{Kohno1DHeisH}. 
In fact, signatures of the high-energy continuum have been observed in inelastic neutron scattering experiments 
on the 1D Heisenberg antiferromagnet CuCl$_2\cdot$2N(C$_5$D$_5$) in a magnetic field \cite{CPC}. 
This high-energy continuum cannot be explained in terms of either magnons in linear spin-wave theory or $\psi$'s and $\psi^*$'s. 
By using the Bethe ansatz \cite{Bethe}, the continuum has been identified \cite{Kohno1DHeisH} as that of 2-string solutions \cite{Takahashi1DHeis}, 
and a QP representing the 2-string has been introduced \cite{Kohno1DHeisH}. 
Such a high-energy feature is not special to 1D chains, but is also relevant to frustrated antiferromagnets in higher dimensions. 
In anisotropic-2D frustrated Heisenberg antiferromagnets, a high-energy mode, which can be regarded as bound states of $\psi$ and the QP for the 2-string, 
emerges from the continuum of 1D chains \cite{Kohno2DHeisH}. Indeed, signatures of the high-energy mode have been observed in inelastic neutron scattering experiments 
on the anisotropic triangular Heisenberg antiferromagnet Cs$_2$CuCl$_4$ in a magnetic field \cite{Coldea}. 
For intuitive understanding of the nature of such high-energy states, it would be helpful to indicate relations to well-known features in other systems. 
In this paper, we argue that the high-energy continuum of antiferromagnetic Heisenberg chains and the high-energy mode of frustrated Heisenberg antiferromagnets in a magnetic field 
are caused by the mechanism similar to that of the upper Hubbard band in the Hubbard model, by using the Bethe ansatz. 

\section{Upper Hubbard band and Doublon}
The Hubbard model is defined by the following Hamiltonian: 
\begin{equation}
{\cal H}_{\rm Hub}=-t\sum_{\langle i,j\rangle\sigma}\left(c^{\dagger}_{i\sigma}c_{j\sigma}+{\rm H.c.}\right)
+U\sum_{i}n_{i\uparrow}n_{i\downarrow}-\mu\sum_{i\sigma}n_{i\sigma}, 
\label{eq:Hub}
\end{equation}
where $c_{i\sigma}$ and $n_{i\sigma}$ denote annihilation and number operators of an electron with spin $\sigma$ at site $i$, respectively. 
Here, $\langle i,j \rangle$ means that $i$ and $j$ are nearest neighbors. 
In the repulsive ($U>0$) Hubbard model, the ground state at half-filling becomes insulating 
when the on-site Coulomb repulsion is strong enough, because addition of an electron to the ground state costs energy of order $U$. 
This implies that the band splits into two bands, called Hubbard bands, by the repulsive interaction. 
One might naively think that double occupancy would characterize the upper Hubbard band and that it may be regarded as a QP, doublon. 
Here, we show that this naive picture is not accurate, by using the 1D Hubbard model. 
In the Bethe ansatz, exact solutions of the 1D Hubbard model are parameterized 
by quasimomenta $\{k_j\}$ and rapidities $\{\Lambda_{\alpha}\}$ that satisfy the following Bethe equations \cite{LiebWu}:
\begin{eqnarray*} 
&&Lk_j=2\pi I_j+2\sum_{\alpha=1}^M\tan^{-1}\frac{4t(\Lambda_{\alpha}-\sin k_j)}{U} \quad \mbox{for} \quad j=1 \sim N,\\
&&\sum_{j=1}^N\tan^{-1}\frac{4t(\Lambda_{\alpha}-\sin k_j)}{U}=\pi J_{\alpha}+\sum_{\beta=1}^M\tan^{-1}\frac{2t(\Lambda_{\alpha}-\Lambda_{\beta})}{U} 
\quad \mbox{for} \quad \alpha=1 \sim M,
\label{eq:Hub}
\end{eqnarray*}
where $L$, $N$, and $M$ denote the number of sites, electrons, and down spins, respectively. 
Here, $I_j$ and $J_{\alpha}$ are called Bethe quantum numbers, which are integers or half-odd integers constrained 
by $L$, $N$, and $M$ under the (anti-)periodic boundary condition. 
The energy and momentum of an eigenstate are expressed as $E=-2t\sum_j\cos k_j-\mu N$ and $K=\sum_j k_j$, respectively. 
Since $\{k_j\}$ and $\{\Lambda_{\alpha}\}$ are obtained through the above Bethe equations once $\{I_j\}$ and $\{J_{\alpha}\}$ are given, 
eigenstates are characterized by $\{I_j\}$ and $\{J_{\alpha}\}$. Essential spectral features in the lower Hubbard band have been well explained 
in terms of QP's defined by the distributions of $\{I_j\}$ and $\{J_{\alpha}\}$ for real $\{k_j\}$ and real $\{\Lambda_{\alpha}\}$ solutions \cite{Benthien1DHub,Kohno1DHub}. 
Here, the QP's defined by $\{I_j\}$ are called holons and anti-holons, and those defined by $\{J_{\alpha}\}$ are called spinons. 
In the lower Hubbard band, there are also continua of 2-$\Lambda$-string solutions \cite{Kohno1DHub,Takahashi1DHub}, 
for which the QP representing the 2-$\Lambda$-string \cite{Kohno1DHeisH} is relevant. 
However, for the upper Hubbard band, we need to consider other solutions. 
\par
\begin{figure}
\begin{center}
\includegraphics[width=16cm]{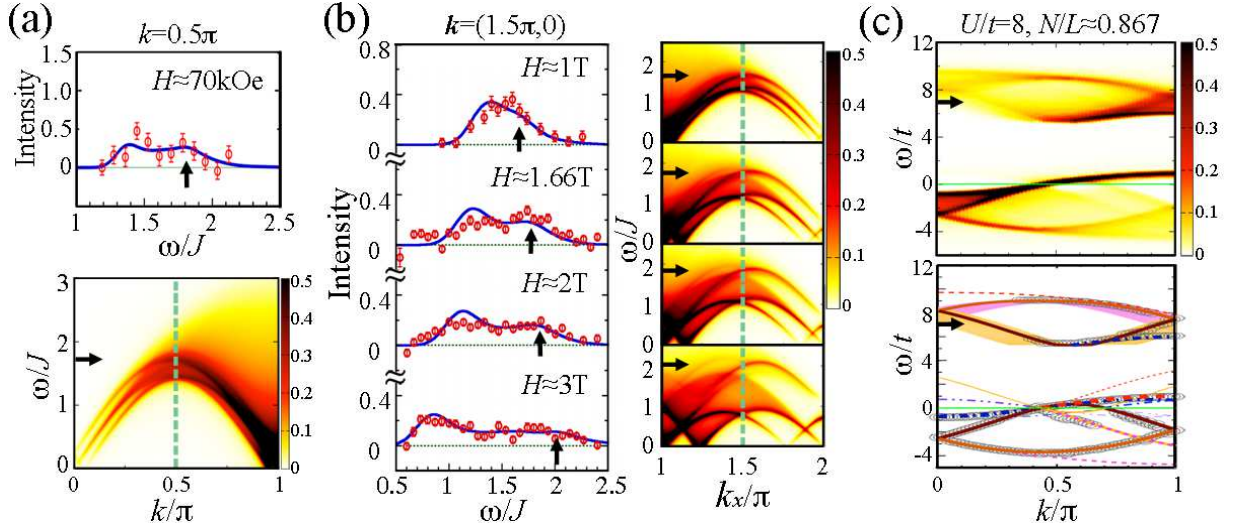}
\end{center}
\caption{(a) Dynamical structure factor $S(k,\omega)=\left[S^{-+}(k,\omega)+S^{+-}(k,\omega)+4S^{zz}(k,\omega)\right]/6$ 
of a spin-1/2 antiferromagnetic Heisenberg chain in a magnetic field. 
The blue solid line and intensity plot show the results calculated using the Bethe ansatz in $L=320$ in Ref. \cite{Kohno1DHeisH}. 
Red open circles denote the experimental results on the 1D Heisenberg antiferromagnet CuCl$_2\cdot$2N(C$_5$D$_5$) in Ref. \cite{CPC}. 
Arrows indicate the high-energy states of the 2-string solutions. 
(b) $S(\mbox{\boldmath$k$},\omega)$ of a spin-1/2 anisotropic triangular Heisenberg antiferromagnet in a magnetic field. 
Blue solid lines and intensity plots show the results calculated using a weak-interchain coupling approximation combined with Bethe-ansatz results of 1D chains in Ref. \cite{Kohno2DHeisH}. 
The ratio of the interchain coupling constant $J^{\prime}$ to the intrachain coupling constant $J$ was chosen as $J^{\prime}/J=0.34$, 
corresponding to that of the anisotropic triangular Heisenberg antiferromagnet Cs$_2$CuCl$_4$ 
($J=0.374$ meV and $J^{\prime}=0.128$ meV \cite{Cs2CuCl4}). 
Red open circles denote the experimental results on Cs$_2$CuCl$_4$ in Ref. \cite{Coldea}. 
Arrows indicate the high-energy mode originating from the 2-string solutions. 
(c) The upper panel shows the single-particle spectral function $A(k,\omega)$ of a repulsive Hubbard chain for $U/t=8$ at $N/L\simeq 0.867$ 
calculated using the dynamical density-matrix renormalization group method in $L=60$ in Ref. \cite{Kohno1DHub}. 
The lower panel shows dispersion relations calculated using the Bethe ansatz in $L=240$ in Ref. \cite{Kohno1DHub}. 
Solid lines show holon modes and anti-holon modes. Blue dashed-dotted lines denote spinon modes. 
Purple dashed double-dotted lines indicate modes of 2-$\Lambda$-string solutions. 
Dashed lines denote edges of continua. The lines in the upper Hubbard band are those in the $k$-$\Lambda$ string solutions. 
Pink and yellow regions represent doublon-holon continua. Open circles indicate peaks of dominant modes of the upper panel. 
Arrows indicate the upper Hubbard band, i.e., the high-energy states of the $k$-$\Lambda$ string solutions.}
\label{fig}
\end{figure}

It is known that there are solutions with complex $\{k_j\}$ \cite{Takahashi1DHub}. 
Recently, single-particle properties in the upper Hubbard band have been explained by using $k$-$\Lambda$ string solutions \cite{Kohno1DHub} 
where the $k$-$\Lambda$ string is a pair of quasimomenta $k^+$ and $k^-$ complex conjugate to each other, 
i.e., $k^{\pm}=\pi-\sin^{-1}\left(\Lambda^{\prime}\pm\frac{\i U}{4t}\right)$ with real $\Lambda^{\prime}$ within the accuracy of $O(\e^{-cL})$ with $c>0$ \cite{Takahashi1DHub}. 
In the $k$-$\Lambda$ string solutions, $\{k_j\}$, $\{\Lambda_{\alpha}\}$, and $\Lambda^{\prime}$ satisfy the following Bethe-Takahashi equations \cite{Takahashi1DHub}:
\begin{eqnarray*} 
&&L\mbox{Re}\left[\sin^{-1}\left(\Lambda^{\prime}+\frac{\i U}{4t}\right)\right]=-\pi J^{\prime}_1+\sum_{j=1}^{N-2}\tan^{-1}\frac{4t(\Lambda^{\prime}-\sin k_j)}{U},\\
&&Lk_j=2\pi I_j+2\tan^{-1}\frac{4t(\Lambda^{\prime}-\sin k_j)}{U}+2\sum_{\alpha=1}^{M-1}\tan^{-1}\frac{4t(\Lambda_{\alpha}-\sin k_j)}{U} \quad \mbox{for} \quad j=1 \sim N-2,\\
&&\sum_{j=1}^{N-2}\tan^{-1}\frac{4t(\Lambda_{\alpha}-\sin k_j)}{U}=\pi J_{\alpha}+\sum_{\beta=1}^{M-1}\tan^{-1}\frac{2t(\Lambda_{\alpha}-\Lambda_{\beta})}{U} 
\quad \mbox{for} \quad \alpha=1 \sim M-1.
\end{eqnarray*}
To explain the properties in the upper Hubbard band, a QP representing the $k$-$\Lambda$ string has been defined 
by using the Bethe quantum number for the $k$-$\Lambda$ string, $J^{\prime}_1$ \cite{Kohno1DHub}. 
Noting that the $k$-$\Lambda$ string has energy of order $U$ in the large $U/t$ regime, i.e., $E_{k-\Lambda}=\mbox{Re}\sqrt{16t^2-(4t\Lambda^{\prime}-\i U)^2}$ \cite{Takahashi1DHub} 
and that it represents a bound state of electrons \cite{Takahashi1DHub}, we can regard the QP as a doublon \cite{Kohno1DHub}. 
This doublon characterizes the upper Hubbard band: while properties in the lower Hubbard band are explained by using solutions without $k$-$\Lambda$ strings, 
those in the upper Hubbard band are explained by using solutions with the $k$-$\Lambda$ string (Fig. \ref{fig}(c)) \cite{Kohno1DHub,multi-string}. 
Also, it should be noted that the number of the doublons is an integer in each eigenstate. 
These features contrast with conventional doublons defined as double occupancies. Namely, the number of double occupancies is not an integer in eigenstates, and double occupancy 
exists even in the ground state at half-filling, i.e., in the lower Hubbard band, which implies that double occupancies do not behave as QP's for the upper Hubbard band. 
In terms of the doublon defined as the $k$-$\Lambda$ string, the upper Hubbard band can be well distinguished from the lower Hubbard band 
by the presence of the doublon ($k$-$\Lambda$ string) in eigenstates. 
Or, we can define the upper and lower Hubbard bands as eigenstates with and without the doublon, respectively. 
From the above argument, we can readily conclude that properties in the lower Hubbard band have no relation to the doublon. 
For example, pseudo-gap behaviors in the lower Hubbard band can be essentially explained in terms of spinon, holons, and anti-holons without the doublon \cite{Kohno1DHub}, 
which contrasts with the picture of doublon (double occupancy)-holon (empty site) 
binding for pseudo-gap behaviors near the Mott transition proposed in the literature \cite{PhilipsRMP,ImadaPRL}. 

\section{High-energy continuum of spin chains in a magnetic field}
The spin-1/2 Heisenberg model is defined by the following Hamiltonian: 
\begin{equation}
{\cal H}_{\rm Heis}=J\sum_{\langle i,j\rangle}\mbox{\boldmath$S$}_i\cdot\mbox{\boldmath$S$}_j-H\sum_{i}S^z_i, 
\label{eq:Heis}
\end{equation}
where $\mbox{\boldmath$S$}_i$ and $S^z_i$ denote the spin-1/2 operator at site $i$ and its $z$-component, respectively. 
In 1D, exact solutions have been obtained by using the Bethe ansatz \cite{Bethe}. 
In the Bethe ansatz, exact solutions of the 1D antiferromagnetic ($J>0$) Heisenberg model are parameterized by rapidities $\{\Lambda_{\alpha}\}$ 
that satisfy the following Bethe equations \cite{YangYang,Bethe}: 
$$
L\tan^{-1}\Lambda_{\alpha}=\pi J_{\alpha}+\sum_{\beta=1}^M\tan^{-1}\frac{\Lambda_{\alpha}-\Lambda_{\beta}}{2} \quad \mbox{for} \quad \alpha=1 \sim M, 
$$
where $J_{\alpha}$ are Bethe quantum numbers having integers or half-odd integers constrained by $L$ and $M$ under the periodic boundary condition. 
In a magnetic field, dominant low-energy properties have been explained in terms of $\psi$'s and $\psi^*$'s \cite{KarbachPsinon,Kohno1DHeisH} 
which are defined as holes and particles created from the consecutive distribution of $\{J_{\alpha}\}$ in the ground state in a magnetic field, respectively \cite{KarbachPsinon}. 
The $\psi$'s reduce to spinons in the zero-field limit. 
In addition to the low-energy continua of $\psi$'s and $\psi^*$'s, a high-energy continuum appears in a magnetic field \cite{Kohno1DHeisH} in the dynamical structure factor 
$S^{+-}(k,\omega)$. Here, dynamical structure factors are defined as $S^{{\bar \alpha}\alpha}(k,\omega)=\sum_i|\langle i|S^{\alpha}_k|{\rm GS}\rangle|^2\delta(\omega-\epsilon_i)$ 
for $\alpha=+,-$, and $z$, where $|i\rangle$ and $|{\rm GS}\rangle$ denote an excited state with excitation energy $\epsilon_i$ and the ground state, respectively. 
The high-energy continuum has been identified as that of 2-string solutions \cite{Kohno1DHeisH} where the 2-string is a pair of rapidities $\Lambda^+$ and $\Lambda^-$ 
complex conjugate to each other, i.e., $\Lambda^{\pm}={\bar \Lambda}\pm\i$ with real ${\bar \Lambda}$ within the accuracy of $O(\e^{-cL})$ with $c>0$ \cite{Takahashi1DHeis}. 
In the 2-string solutions, $\{\Lambda_{\alpha}\}$ and ${\bar \Lambda}$ satisfy the following Bethe-Takahashi equations \cite{Takahashi1DHeis}: 
\begin{eqnarray*}
&&L\tan^{-1}\frac{{\bar \Lambda}}{2}=\pi {\bar J}_1
+\sum_{\beta=1}^{M-2}\left[\tan^{-1}\frac{{\bar \Lambda}-\Lambda_{\beta}}{3}+\tan^{-1}\left({\bar \Lambda}-\Lambda_{\beta}\right)\right],\\
&&L\tan^{-1}\Lambda_{\alpha}=\pi J_{\alpha}+\tan^{-1}\frac{\Lambda_{\alpha}-{\bar \Lambda}}{3}+\tan^{-1}\left(\Lambda_{\alpha}-{\bar \Lambda}\right)
+\sum_{\beta=1}^{M-2}\tan^{-1}\frac{\Lambda_{\alpha}-\Lambda_{\beta}}{2}, 
\end{eqnarray*}
for $\alpha=1 \sim M-2$. By using the Bethe quantum number for the 2-string, ${\bar J}_1$, a QP representing the 2-string has been defined \cite{Kohno1DHeisH}, 
which can be regarded as a bound state of two $\psi^*$'s \cite{Kohno1DHeisH}. 
Thus, the 2-string in the Heisenberg model is analogous to the $k$-$\Lambda$ string in the Hubbard model: 
the high-energy states in the Heisenberg model in a magnetic field and those in the Hubbard model are both due to string solutions with string length greater than one. 
In other words, the QP responsible for the high-energy continuum in the Heisenberg model in a magnetic field is related in origin with the doublon in the Hubbard model through the Bethe ansatz. 
\par
The physical picture of this correspondence relation will be more intuitively understood by mapping the spin-1/2 antiferromagnetic Heisenberg model to the hard-core boson model with nearest neighbor repulsion:
\begin{equation}
{\cal H}_{\rm hcb}=\frac{J_{xy}}{2}\sum_{\langle i,j\rangle}\left(b^{\dagger}_ib_j+{\rm H.c.}\right)
+J_z\sum_{\langle i,j\rangle}\left(n_{{\rm b}i}-\frac{1}{2}\right)\left(n_{{\rm b}j}-\frac{1}{2}\right)+H\sum_{i}\left(n_{{\rm b}i}-\frac{1}{2}\right), 
\label{eq:b}
\end{equation}
where $b_i$ and $n_{{\rm b}i}$ denote annihilation and number operators of a hard-core boson at site $i$. For the Heisenberg model, $J_{xy}=J_z=J$. 
As in the Hubbard model, the repulsive interaction ($J_z$-term) raises energy, when two bosons are located at neighboring sites. 
Thus, it is naturally expected that the band will split due to the repulsion, if the repulsive interaction is strong enough ($J_z\gg J_{xy}$). 
For the Heisenberg model ($J_{xy}=J_z$), although the band does not split in zero field, it splits in a magnetic field \cite{Kohno1DHeisH}. 
The continuum of 2-string solutions shifts to higher energies as the magnetic field increases \cite{Kohno1DHeisH}, 
as the upper Hubbard band does with hole doping \cite{Kohno1DHub}. 
Likewise, a corresponding continuum appears in the spinless fermion model with nearest neighbor repulsion on a 1D lattice \cite{Kohno1DSF,Pereira1DSF}, 
since the hard-core boson model can be mapped to the spinless fermion model through the Jordan-Wigner transformation. 

\section{High-energy mode of frustrated antiferromagnets in a magnetic field}
In anisotropic-2D frustrated Heisenberg antiferromagnets, interchain spin exchange induces bound states of 1D QP's from continua of 1D chains \cite{Kohno2DHeis,Kohno2DHeisH}. 
The mechanism can be intuitively understood by noting that spin exchange corresponds to two-spinon hopping in zero field \cite{Kohno2DHeis}: 
the kinetic energy in the direction perpendicular to chains gains by forming bound states of spinons due to the pair hopping of spinons, i.e., spin exchange, between neighboring chains. 
The modes of such bound states dominate the spectra and behave as QP's in anisotropic-2D frustrated Heisenberg antiferromagnets \cite{Kohno2DHeis,Kohno2DHeisH}. 
This implies that the origin of the dominant modes can be traced back to the 1D QP's. 
Namely, the high-energy mode in the frustrated Heisenberg antiferromagnets in a magnetic field originates from $\psi$ and the QP for the 2-string \cite{Kohno2DHeisH}. 
Thus, the high-energy mode observed in the anisotropic triangular Heisenberg antiferromagnet Cs$_2$CuCl$_4$ in a magnetic field \cite{Coldea} can be interpreted as the mode 
originating from the mechanism similar to that of the upper Hubbard band in a Hubbard chain. 

\section{Summary}
In spin-1/2 antiferromagnetic Heisenberg chains and 
anisotropic-2D frustrated Heisenberg antiferromagnets, high-energy states carry considerable spectral weights in a magnetic field  \cite{Kohno1DHeisH,Kohno2DHeisH,CPC,Coldea}. 
Such high-energy properties cannot be explained in terms of either Nambu-Goldstone bosons due to spontaneous breaking of continuous symmetries 
or gapless QPs in a Tomonaga-Luttinger liquid. 
In this paper, the origin of the high-energy states in the spin-1/2 Heisenberg antiferromagnets in a magnetic field was shown to be analogous 
to that of the upper Hubbard band in the 1D repulsive Hubbard model, 
by using the Bethe ansatz: the high-energy states in both the Heisenberg antiferromagnets and the Hubbard model originate from string solutions with string length greater than one, 
i.e., the 2-string solutions for the Heisenberg antiferromagnets and the $k$-$\Lambda$ string solutions for the Hubbard model. 
Thus, we can interpret the high-energy states in spin-1/2 1D Heisenberg antiferromagnets (Fig. \ref{fig}(a)) and anisotropic-2D frustrated Heisenberg antiferromagnets (Fig. \ref{fig}(b)) 
observed in inelastic neutron scattering experiments in a magnetic field \cite{CPC,Coldea} 
as those caused by the mechanism similar to that of the upper Hubbard band in the Hubbard model (Fig. \ref{fig}(c)). 
In other words, the QP's responsible for the high-energy states in the Heisenberg antiferromagnets in a magnetic field are related in origin with the doublon in the Hubbard model through the Bethe ansatz. 
It should be noted that, in contrast to an anti-bonding band due to multi-site unit cells, the high-energy states are induced by repulsive interactions without sublattices. 
Thus, the high-energy properties of the Heisenberg antiferromagnets in a magnetic field can be understood in the context of the Mott physics. 
I expect that such high-energy states will generally appear in the one-particle addition spectral function in repulsively interacting systems if the number of particles is small enough. 

\section*{Acknowledgments}
This work was supported by 
World Premier International Research Center Initiative on Materials Nanoarchitectonics, 
MEXT, Japan, and KAKENHI 22014015.


\begin{thebibliography}{9}
\bibitem{multi-string} There are solutions with multiple $k$-$\Lambda$ strings and longer $k$-$\Lambda$ strings \cite{Takahashi1DHub}. 
But, they have much higher energies \cite{Takahashi1DHub} and much smaller single-particle spectral weights \cite{Kohno1DHub} 
than the solutions with the $k$-$\Lambda$ string defined as a pair of quasimomenta in the large $U/t$ regime.
\end{thebibliography}
\end{document}